\documentclass[fleqn,usenatbib]{mnras}

\usepackage{mathptmx}

 \usepackage[T1]{fontenc}

\usepackage{graphicx}	
\usepackage{amsmath}	
\usepackage{amssymb}	
\usepackage[usenames,dvipsnames]{xcolor} 
\usepackage[normalem]{ulem} 

\newcommand{\OIIIHb}{\mbox{[O\textsc{iii}]+H$\beta$}}

\newcommand{\synthesizer}{\mbox{\textsc{synthesizer}}}

\newcommand{\jwst}{\mbox{\it JWST}}

\title[CEERS: Colour Evolution]{Cosmic Evolution Early Release Science (CEERS) survey: The colour evolution of galaxies in the distant Universe}

\author[Stephen M. Wilkins et al.]{Stephen M. Wilkins$^{1,2}$\thanks{E-mail: s.wilkins@sussex.ac.uk}, 
Jack C. Turner$^{1}$, 
Micaela B. Bagley$^{3}$,
Steven L. Finkelstein$^{3}$,
Ricardo O. Amor\'{i}n$^{4,5}$, \newauthor
Adrien Aufan Stoffels D Hautefort$^{1}$, 
Peter Behroozi$^{6,7}$,
Rachana Bhatawdekar$^{8}$,
Avishai Dekel$^{9}$, \newauthor
James Donnellan$^{1}$, 
Nicole E. Drakos$^{10}$, 
Flaminia Fortuni$^{11}$, 
Nimish P. Hathi$^{12}$,
Michaela Hirschmann$^{13}$, \newauthor
B. W. Holwerda$^{14}$, 
Dimitrios Irodotou$^{15}$ 
Anton M. Koekemoer$^{12}$, 
Christopher C. Lovell$^{16,1}$,
Emiliano Merlin $^{11}$,  \newauthor
Will J. Roper$^{1}$, 
Louise T. C. Seeyave$^{1}$, 
Aswin P. Vijayan $^{17,18}$, and
L. Y. Aaron {Yung}$^{19,12}$
\\
$^{1}$Astronomy Centre, University of Sussex, Falmer, Brighton BN1 9QH, UK\\
$^{2}$Institute of Space Sciences and Astronomy, University of Malta, Msida MSD 2080, Malta \\
$^{3}$Department of Astronomy, The University of Texas at Austin, Austin, TX, USA\\
$^{4}$ARAID Foundation. Centro de Estudios de F\'{\i}sica del Cosmos de Arag\'{o}n (CEFCA), Unidad Asociada al CSIC, Plaza San Juan 1, E--44001 Teruel, Spain\\
$^{5}$ Departamento de Astronom\'{i}a, Universidad de La Serena, Av. Juan Cisternas 1200 Norte, La Serena 1720236, Chile\\
$^{6}$Department of Astronomy and Steward Observatory, University of Arizona, Tucson, AZ 85721, USA\\
$^{7}$Division of Science, National Astronomical Observatory of Japan, 2-21-1 Osawa, Mitaka, Tokyo 181-8588, Japan\\
$^{8}$European Space Agency (ESA), European Space Astronomy Centre (ESAC), Camino Bajo del Castillo s/n, 28692 Villanueva de la Cañada, Madrid, Spain\\
$^{9}$Racah Institute of Physics, The Hebrew University, Jerusalem 91904, Israel\\
$^{10}$Department of Physics and Astronomy, University of Hawaii, Hilo, 200 W Kawili St, Hilo, HI 96720, USA\\
$^{11}$ INAF - Osservatorio Astronomico di Roma, via Frascati 33, 00078 Monte Porzio Catone (Roma), Italy\\
$^{12}$ Space Telescope Science Institute, 3700 San Martin Drive, Baltimore, MD 21218, USA\\
$^{13}$ Institute of Physics, Laboratory of Galaxy Evolution, Ecole Polytechnique Fédérale de Lausanne (EPFL), Observatoire de Sauverny, 1290 Versoix, Switzerland\\
$^{14}$ Department of Physics and Astronomy, University of Louisville, Natural Science Building 102, 40292 KY, Louisville, USA\\
$^{15}$Department of Physics, University of Helsinki, Gustaf Hällströmin katu 2, FI-00014, Helsinki, Finland\\
$^{16}$Institute of Cosmology \& Gravitation, University of Portsmouth, Dennis Sciama Building, Portsmouth, PO1 3FX, UK\\
$^{17}$Cosmic Dawn Center (DAWN)\\
$^{18}$DTU-Space, National Space Institute, Technical University of Denmark, Elektrovej 327, 2800 Kgs.~Lyngby, Denmark\\
$^{19}$Astrophysics Science Division, NASA Goddard Space Flight Center, 8800 Greenbelt Rd, Greenbelt, MD 20771, USA\\
}

\date{Accepted XXX. Received YYY; in original form ZZZ}

\pubyear{2022}

\begin{document}
\label{firstpage}
\pagerange{\pageref{firstpage}--\pageref{lastpage}}
\maketitle

\begin{abstract}
The wavelength-coverage and sensitivity of \jwst\ now enables us to probe the rest-frame UV - optical spectral energy distributions (SEDs) of galaxies at high-redshift ($z>4$). From these SEDs it is, in principle, through SED fitting possible to infer key physical properties, including stellar masses, star formation rates, and dust attenuation. These in turn can be compared with the predictions of galaxy formation simulations allowing us to validate and refine the incorporated physics. However, the inference of physical properties, particularly from photometry alone, can lead to large uncertainties and potential biases. Instead, it is now possible, and common, for simulations to be  \emph{forward-modelled} to yield synthetic observations that can be compared directly to real observations. In this work, we measure the \jwst\ broadband fluxes and colours of a robust sample of $5<z<10$ galaxies using the Cosmic Evolution Early Release Science (CEERS) Survey. We then analyse predictions from a variety of models using the same methodology and compare the NIRCam/F277W magnitude distribution and NIRCam colours with observations. We find that the predicted and observed magnitude distributions are similar, at least at $5<z<8$. At $z>8$ the distributions differ somewhat, though our observed sample size is small and thus susceptible to statistical fluctuations. Likewise, the predicted and observed colour evolution show broad agreement, at least at $5<z<8$. There is however some disagreement between the observed and modelled strength of the strong line contribution. In particular all the models fails to reproduce the F410M-F444W colour at $z>8$, though, again, the sample size is small here.
\end{abstract}

\begin{keywords}
	galaxies: general -- galaxies: evolution -- galaxies: formation -- galaxies: high-redshift -- galaxies: photometry 
\end{keywords}



\section{Introduction}\label{sec:intro}

A key objective in extragalactic astrophysics is to constrain the physical processes responsible for galaxy formation and evolution. These include the accretion and cooling of gas onto and in galaxies, star formation, super-massive black hole formation and growth, and feedback from stars and AGN, including metal enrichment and dust creation and destruction. These physical processes ultimately manifest in the observable spectral energy distributions (SEDs) and structure of galaxies. Thus, by comparing observations with models it becomes possible to constrain these physical processes.

Overwhelmingly, the most common method is to utilise spectral energy distribution SED fitting \citep[see][for an overview]{Conroy2013, Pacifici23} to use photometric (and spectroscopic) observations to constrain key physical properties such as stellar masses, star formation histories, dust attenuation, and metallicities. These measurements can, in turn, be compared \emph{directly} with galaxy formation model predictions for integrated galaxy properties and their cosmological distribution functions, and possibly used to constrain uncertain parameters in those models \citep[e.g][]{crain_eagle_2015}. 

However, SED fitting can yield large uncertainties for individual galaxies and can result in complex biases \citep[see e.g.][]{Conroy2009, Pacifici2015, Carnall2019, Lower2020, Meldorf23, Pacifici23}. As an example, in some scenarios, strong optical line emission, indicative of recent star formation, can be mistaken for a strong Balmer/4000\AA\ break feature and thus an evolved galaxy. This, in turn, can result in dramatic overestimates of the stellar mass due to the large difference in mass-to-light ratios between young and evolved galaxies \citep[e.g.][]{Endsley23}.

In addition, when comparing populations, for example between observed and theoretical samples, it is essential to understand the sample \emph{completeness}. For example, when measuring a distribution function (e.g. the far-UV luminosity function, or galaxy stellar mass function) it is necessary to correct for incompleteness, particularly toward the sensitivity limit. In practice, this requires making some assumptions about the morphologies and SEDs of the real sources and testing the recoverability of inserted sources \citep[see e.g.][]{GLACiAR}. While these assumptions can be motivated by observational constraints, for example employing constraints from deeper observations, to motivate the assumed properties of the source, this inevitably introduces an additional and complex source of bias.  

To avoid both these issues, an alternative approach is to directly compare observed SEDs with synthetic observations generated from galaxy formation models. This is now increasingly possible thanks to sophisticated, and fast, forward-modelling pipelines that are capable of producing a range of synthetic observations \citep[e.g.][]{Blaizot2005, Wilkins13, Snyder15, Wilkins16, Laigle2019, Bravo2020, FLARES-II, Fortuni23, Snyder23}. This not only allows us to avoid some of the uncertainties and biases introduced by SED fitting but also avoids the need for complicated and uncertain completeness corrections, assuming that galaxies are selected with identical criteria. 


In this work, we adopt this approach to study the evolution of galaxy populations at high-redshift ($z>5$) by directly comparing observations of galaxies with model predictions. In doing so we will provide a more robust test of galaxy formation models and the forward modelling applied to them.

We begin by identifying a sample of galaxies at $5<z<10$ using observations from the Cosmic Evolution Early Release Science (CEERS) survey \citep{Bagley2023}. We then measure their broadband spectral energy distributions, specifically using their NIRCam colours. We then compare the evolution of these colours with a handful of theoretical predictions including the First Light And Reionisation Epoch Simulations \citep[FLARES]{FLARES-I, FLARES-II, FLARES-VI}, the Santa Cruz semi-analytical model \citep[SCSAM]{Somerville2015, Somerville2021, Yung2022}, and the semi-empirical JAGUAR \citep{Williams2018} and DREaM \citep{Drakos2022} models. By doing so we can identify sources of disagreement and thus areas of possible refinement for those models.

This article is organised as follows: we begin, in Section \ref{sec:theory}, by describing some of the theoretical background to galaxy SEDs. We then proceed, in Section \ref{sec:observations}, by describing the observed sample of galaxies including the NIRCam reduction (\S\ref{sec:obs:reduction}), source identification (\S\ref{sec:obs:id}), photometry (\S\ref{sec:obs:photom}), photometric redshift measurement (\S\ref{sec:obs:pz}), and selection criteria (\S\ref{sec:obs:selection}). Next, in Section \ref{sec:models} we describe the four models utilised in our comparison and the additional steps taken to produce a sample aligned with the observed sample. In Section \ref{sec:results} we present our results, including the observed and predicted F277W flux distribution (\S\ref{sec:results:magnitude}) and colour evolution (\S\ref{sec:results:colours}). Finally, in Section \ref{sec:conclusions}, we present our conclusions and future directions (\S\ref{sec:conclusions:future}).

\section{Theory}\label{sec:theory}

The \emph{intrinsic} spectral energy distributions (SEDs) of galaxies are driven by their star formation and metal enrichment histories and the presence of active galactic nuclei (AGN). These intrinsic SEDs are then modified through reprocessing by gas and dust, notably producing strong nebular line emission and reddening by dust. Observation of the spectral energy distribution then \emph{potentially} allows us to constrain the star formation and metal enrichment history (and thus the stellar mass, star formation rate and history, and metallicity), the contribution of AGN, the dust attenuation, and the escape fraction of Lyman-continuum photons ($f_{\rm esc}$).

\begin{figure}
 	\includegraphics[width=\columnwidth]{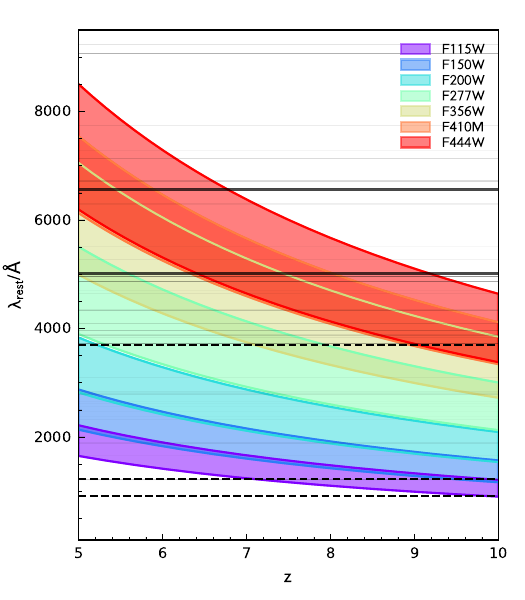}
	\caption{The rest-frame wavelength probed by the seven JWST/NIRCam filters observed by CEERS as a function of redshift. The dashed horizontal lines show (from top to bottom) the location of the Balmer-limit, Lyman-$\alpha$, and Lyman-limit breaks. The solid horizontal lines denote various emission lines with the line-width and opacity scaled by their relative intensity for a young dust-free star forming galaxy. 
	\label{fig:rest}}
\end{figure}

In the context of the CEERS survey we have access to six NIRCam wide filters (F115W, F150W, F200W, F277W, F356W, and F444W), the NIRCam/F410M medium band, and several HST bands. The seven NIRCam filters probe the rest-frame UV-optical at $z=5-10$ as shown in Figure \ref{fig:rest}. This Figure also highlights the location of the Lyman, Lyman-$\alpha$, and Balmer limits, and rest-frame UV and optical emission lines, weighted by their relative intensity for the standard model described below. 

\begin{figure}
 	\includegraphics[width=\columnwidth]{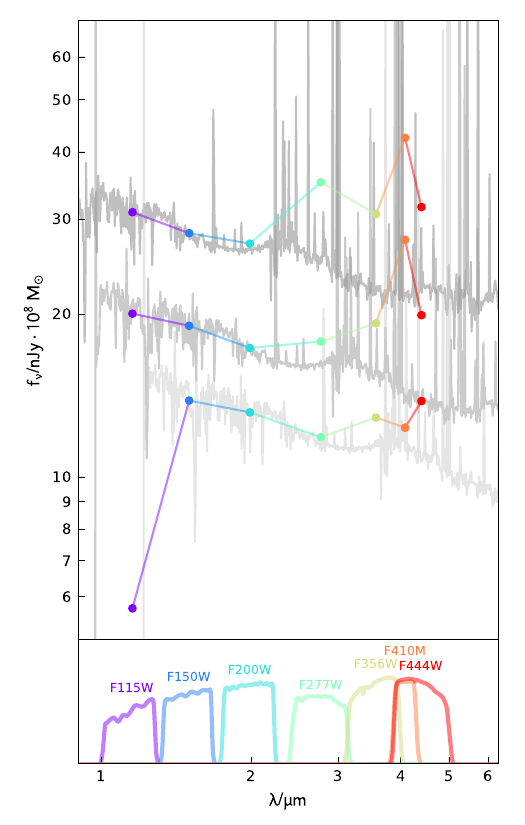}
	\caption{Example model spectra (generated using the \synthesizer\ code, Lovell et al. \emph{in-prep}) of a galaxy that has formed a total of $10^8\ {\rm M_{\odot}}$ of stars with a constant star formation history over the preceding 100 Myr, observed at $z\in\{5,7,9\}$ (top, middle, bottom lines respectively). Coloured points show the expected fluxes in the seven \emph{JWST}/NIRCam filters observed by CEERS. The H$\alpha$ equivalent width of this galaxies is $\approx 600$ \AA. The lower panel shows the filter transmission functions of the seven filters.
	\label{fig:sed}}
\end{figure}

\begin{figure}
 	\includegraphics[width=\columnwidth]{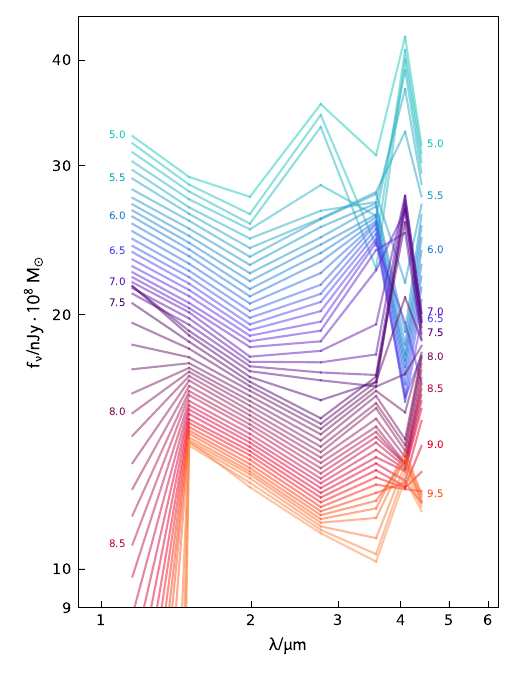}
	\caption{Predicted fluxes for the same model spectra shown in Figure \ref{fig:sed} but showing a finer grid of redshifts.
	\label{fig:sedz}}
\end{figure}

\begin{figure}
 	\includegraphics[width=\columnwidth]{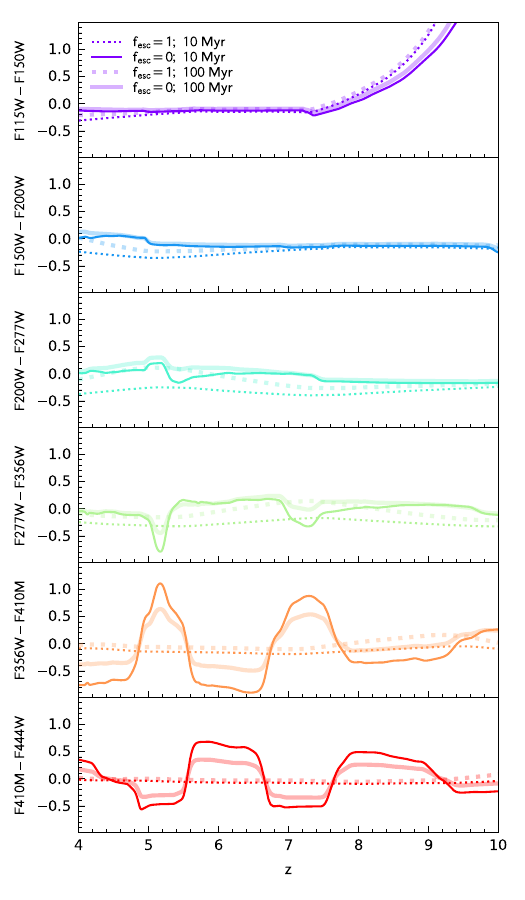}
	\caption{Predicted colour evolution for galaxies with 10 or 100 Myr continuous star formation, $Z_{\star}=0.01$, and both $f_{\rm esc}\in\{0,1\}$. 
	\label{fig:colour_evolution}}
\end{figure}

Synthetic spectra, broadband fluxes, and colours, generated using the \synthesizer\footnote{\url{https://flaresimulations.github.io/synthesizer/}}\ (Lovell et al. \emph{in-prep}) synthetic observations pipeline, for a simple model star-forming galaxies at $z\in\{5,7,9\}$ are shown in Figure \ref{fig:sed}. This model assumes 100 Myr continuous star formation, $Z_{\star}=0.01$, and $f_{\rm esc}=0$ and is generated using the v2.2.1 of the Binary Population And Spectral Synthesis \citep[\textsc{bpass},][]{BPASS2.2.1} stellar population synthesis (SPS) code, assuming a \citet{chabrier_galactic_2003} initial mass function. Nebular (H\textsc{ii} region) emission, including both line and continuum emission, is modelled using the v17.03 of the \textsc{cloudy} photoionisation code \citep{Cloudy17.02} assuming $\log_{10}U = -2$, $n_{e}=100\ {\rm cm^{-3}}$, $Z_{\rm gas}=Z_{\star}$ and spherical geometry. Similarly, Figure \ref{fig:sedz} shows the same model SED at a wider selection of redshifts, but only the broadband fluxes and colours. Figure \ref{fig:colour_evolution} instead shows the evolution of the six adjacent colours based on the CEERS NIRCam filters, but for both a 100 and 10 Myr constant star formation history model with Lyman-continuum escape fractions of $f_{\rm esc}=0$ and $1$ (labelled "pure stellar" and "stellar + nebular" respectively). Collectively these figures demonstrate the expected features and redshift evolution of a star-forming galaxy. Most notable, other than the dimming with redshift, is the effect of line emission. This results in a strong sensitivity of the observable colours to redshift, particularly in the F356W$-$F410M and F410M$-$F444W colours, as various strong lines move in and out of the F410M band. Clearly then a statistical analysis of galaxies across this redshift range can be used to probe the strength of line emission thus providing insights into to the star formation histories of galaxies at this epoch.

In addition to the impact of line emission, Figure \ref{fig:colour_evolution} also shows the impact of changing the duration of previous star formation. Reducing the duration broadly results in colours becoming bluer and the magnitude of colour fluctuations caused by nebular line emission increasing. This is entirely expected as the primary effect of a longer star formation epoch is to boost the contribution of slightly longer-lived, less massive stars relative to the most massive. These contribute strongly to the UV continuum but are typically less strongly ionizing, so contribute less to the nebular emission. As noted above, galaxy colours are also affected by the presence of an AGN, the metallicity, dust attenuation, and more broadly the specific shape of the star formation history. In addition, predicted colours, and thus the physical properties inferred from observed colours, are also sensitive to the choice of SPS model and IMF. Several of these effects are explored in more detail in \citet{FLARES-VI}, using an earlier version of the \synthesizer\ pipeline.

\section{Observations}\label{sec:observations}

CEERS is an Early Release Science program (Proposal ID 1345, PI: Finkelstein) that has now completed its NIRCam, NIRSpec and MIRI observations of the CANDELS Extended Groth Strip (EGS). In this work, we make use of the four NIRCam pointings observed in June 2022 and the remaining six observed the following December, covering a total area of $97\text{ arcmin}^{2}$. Both sets of observations used the F115W, F150W and F200W short-wavelength filters and the F277W, F356W, F410M and F444W bands at long-wavelengths. The reduced images used here are publicly available\footnote{\url{https://ceers.github.io/releases.html}} as data release 0.5 and 0.6 for the June and December pointings respectively. Given that this study bypasses the uncertainties of SED fitting, the effects of assumptions made during the reduction and source extraction process become more prominent. We therefore describe the CEERS approach to both in this section.

\subsection{Reduction}\label{sec:obs:reduction}

A detailed description of the reduction pipeline can be found in \citet{Bagley2023}, but we highlight the key features here. The raw imaging from the June pointings is reduced through version 1.7.2 of the JWST Calibration Pipeline\footnote{\url{https://github.com/spacetelescope/jwst}} \citep{jwstpipeline} with CRDS \textsc{pmap} 0989, while the additional six pointings use the updated pipeline version 1.8.5 and CRDS \textsc{pmap} 1023. 

The steps of the reduction are the same for all ten pointings. After passing through stage one of the pipeline, and creating a count-rate image, custom steps are carried out to correct for additional features. A custom correction is applied to remove snowballs, and large wisps in the F150W and F200W images are removed using NIRCam team templates\footnote{\url{https://stsci.app.box.com/s/1bymvf1lkrqbdn9rnkluzqk30e8o2bne}}. The $1/f$ noise is quantified and removed using a median value measured across rows and columns. Stage 2 of the pipeline performs flat fielding and flux calibration, producing images with units $\text{MJy sr}^{-1}$.

Images are aligned using a custom \textsc{TweakReg} routine which registers an image to an absolute WCS frame by matching sources to a reference catalog. The modified version uses Source Extractor \citep{Bertin1996} to measure source centroids in each image, before aligning these to a reference catalogue constructed based on an HST F160W mosaic of the field from CANDELS \citep{Grogin2011, Koekemoer2011} that had been reprocessed with astrometry tied to Gaia EDR3.

Stage 3 of the pipeline performs an initial background estimation and creates mosaics by drizzling the images to a common output pixel scale of $0.03''\text{pixel}^{-1}$. Any remaining background is estimated and subtracted using a custom Python script. An initial estimate is made using ring-median filtering before four tiers of source masking begin by removing the most extended galaxies moving progressively to the smallest. A final smooth background model is then calculated for the fully masked image.

Finally, the point spread functions (PSFs) of F115W, F150W and F200W are matched to the larger PSF of the longer wavelength F277W band.  For the redder bands (F356W, F410M, F444W) with larger PSFs, correction factors are calculated by convolving the F277W image to the larger PSF, and measuring the flux ratio in the original image to that in the convolved image. A correction factor is then applied in the images with larger PSFs to correct for the missing flux. This approach assumes that the morphology is not significantly affected by the PSF. All PSFs were measured empirically by stacking stars (see \citealt{Finkelstein2023} for details).

\subsection{Source Identification}\label{sec:obs:id}

Source identification and photometry were performed following a similar approach to \citet{Finkelstein2023}, outlined in full by Finkelstein et al. \emph{in-prep}, using \textsc{Source Extractor} v2.25.0 in two-image mode. Sources are identified from a detection image which is the inverse-variance-weighted sum of the PSF-matched F277W and F356W images. Shorter wavelength bands were excluded to avoid missing potential F200W dropout sources, which could be at extremely high-redshift. The NIRCam mosaic for each band is then passed individually as the measurement image. Simulated CEERS imaging \citep{Bagley2023} was initially used to set the detection parameters, but these were then refined by inspecting the outputs and attempting to maximise completeness while minimising spurious sources. The final values of \texttt{DETECT\_THRESH=1.4} and \texttt{DETECT\_MINAREA=5} are the same for all images.

\subsection{Source Photometry}\label{sec:obs:photom}

Photometry is initially measured in small Kron apertures, using parameters tuned to accurately recover colours at high-redshift, before being corrected by two factors. Firstly, \textsc{Source Extractor} is run for a second time on the F277W image using the default aperture parameters which results in larger sizes. The correction is then the ratio between the flux in this larger aperture to the fiducial measurement. A median multiplicative factor of 1.5 was used across all fields.

An additional correction to account for flux missed on larger scales is estimated using source injection simulations. Sersic profiles are injected into the F277W and F356W images using \textsc{GALFIT} \citep{Peg2002}. The F277W+F356W detection image is then regenerated and photometry is measured in F277W. The recovered fluxes are found to be lower by a factor of 1.05 at an apparent magnitude of $m=24$, rising to $1.20$ at $m=28$. A linear relation is fit to the offset and applied to the previously corrected fluxes. As both of these corrections are applied equally to all bands, they affect only the total flux, not the measured colours.

Uncertainties on the corrected fluxes were estimated by placing random non-overlapping apertures of differing size in blank areas of the images. A four-parameter function is then fit to the normalised median absolute deviation measured in each aperture as a function of its area. The flux error for each object is then extracted from this fit based on the size of its Kron aperture.

\subsection{Photometric Redshifts}\label{sec:obs:pz}

Photometric redshifts are estimated for all objects in the catalogue using the Easy and Accurate
$z_{\text{phot}}$ from Yale \citep[\textsc{eazy}]{Brammer2008} template fitting code. For a given source, user-supplied templates are fit in non-negative linear combination to derive a probability distribution function (PDF) for the redshift, which is based on the quality of fit of each possible template combination to the measured photometry. 

The estimates used in this study are derived by the approach adopted by \citet{Finkelstein2023}, using a total of eighteen basis templates. The recommended “tweak\_fsps\_QSF\_12\_v3” set of twelve templates, which are the principal components of a set of 560 synthetic \textsc{fsps} \citep{Conroy2010} spectra, are supplemented with six additional templates generated by \citet{Larson2022}. These were generated by combining stellar population spectra from \textsc{bpass} \citep{BPASS2.2.1} with nebular emission derived with \textsc{cloudy} \citep{Cloudy17.02} and better sample the predicted blue rest-frame UV colours of $z>8$ galaxies. 

While the distribution of bright galaxies is likely to be skewed towards lower redshift, the bright end of the high-redshift luminosity function is yet to be well constrained. Therefore, in order to minimise bias against the selection of true, bright distant galaxies, a flat prior in luminosity is assumed. To account for potentially unknown systematics, a minimum error of 5\% is assumed for each measured flux and these measurements are then fit in an \textsc{eazy} run assuming a flat redshift prior from 0.01-20.

\subsection{Object Selection}\label{sec:obs:selection}

As the CEERS catalogue still contains a number of spurious sources, we next apply a series of cuts which select a robust sample of galaxies at $z=5-10$. Firstly, to ensure that the galaxies are confidently detected, we impose a minimum flux of 50 nJy ($m\approx 27.25$) in the F277W filter, which is part of the detection image. Doing so also aligns us with the depths accessible to all the models; these depend on the resolution of the particular simulation. Additionally, we require a minimum signal-to-noise ratio of $5.5$, as measured in 0.2'' diameter circular apertures, in at least four of the seven NIRCam bands. The remaining galaxies have reliable photometric information across the SED.

To select galaxies within the desired redshift range we apply a further four cuts. The maximum likelihood \textsc{eazy} photometric redshift estimate must satisfy $z_{a}>4.5$ and have an acceptable goodness-of-fit $\chi^{2} <60$, with the latter condition chosen to align with \citet{Finkelstein2023}. To remove objects with potential low-redshift solutions, we require that $\int P(z>4) \geq 0.9$, ensuring that at least 90\% of the posterior probability is above $z=4$. Finally, we require a signal-to-noise ratio $<2$ in bands below the wavelength of the Lyman-$\alpha$ break at the maximum likelihood redshift estimate. If the galaxy is truly at that redshift, it should be undetected in those bands. 

These six cuts are reasonably conservative and should result in our sample of 1112 galaxy candidates at $z>4.5$ being robust. In addition to these cuts, we visually inspect the SED, redshift PDF, segmentation map, detection image and individual cutouts of each galaxy to identify and remove any remaining spurious sources such as diffraction spikes or clear defects in the images. A further 48 objects are removed during this process, resulting in a final sample of 1064. Additionally, any objects that could have their photometry inflated by nearby objects of comparable physical extent or flux are flagged, so any potential systematic effects may be identified.

\section{Models}\label{sec:models}

\subsection{Models considered}\label{sec:models.models}

In this work we compare our observational measurements against four models that provide predictions for the HST and JWST fluxes we are able to measure. These models range in complexity from the semi-empirical to fully hydrodynamical models. Table \ref{tab:model_summary} provides a summary of the four models considered in this work including some of their modelling assumptions.

\begin{table*}
\begin{tabular}{lllllll}
\hline
 Model & Reference(s) & Type & SED modelling & SPS & Nebular emission  \\
\hline\hline
JAGUAR          & \citet{Williams2018}  & Semi-empirical    & \citet{Chevallard2016}        & BC03          & \citet{Chevallard2016} \\
DREaM           & \citet{Drakos2022}    & SHAM              & \emph{internal}                           & FSPS          & \citet{Byler2017} \\
SCSAM           & \citet{Somerville2015}; \citet{Yung2019} & SAM               & \emph{internal}                                & BC03          & \citet{Hirschmann2017, Hirschmann2019, Hirschmann2022}  \\
FLARES          & \citet{FLARES-I}      & Hydro             & \citet{FLARES-II} & BPASS-2.2.1  & \citet{FLARES-II} \\
\hline
\end{tabular}
\caption{Summary of the models considered in this work including the modelling approach, SED modelling methodology, assumed SPS model, and nebular emission modelling methodology. Where an entry is marked \emph{internal} the requisite modelling is described in reference article(s).}\label{tab:model_summary}
\end{table*}

\subsubsection{JAGUAR} \label{sec:JAGUAR}

JAGUAR (JAdes extraGalactic Ultradeep Artificial Realizations)  \citep{Williams2018} is a semi-empirical model describing the evolution of galaxy number counts and spectral energy distributions. JAGUAR adopts different approaches depending on the redshift and luminosity of the source; here we describe the modelling of sources at $z>4$. In this regime JAGUAR utilises a model of the evolution of the galaxy stellar mass function, based on observations of the far-UV luminosity function. This model is then combined with an empirical model linking the stellar mass to the UV luminosity, redshift, and the UV continuum slope $\beta$. Based on these values a spectral energy distribution is assigned using the \textsc{BEAGLE} \citep{Chevallard2016} tool providing a self-consistent treatment of stellar and photo-ionised gas emission and dust attenuation. This modelling assumes the updated version of the \citep{BC03} stellar population synthesis model with photo-ionisation modelling using version 13.3 of the \textsc{cloudy} photoionisation model \citep{cloudy13}. Dust attenuation is accounted for using the \citet{Charlot_and_Fall} two-component model. 

\subsubsection{DREaM}

The Deep Realistic Extragalactic (DREaM) simulated galaxy catalogs \citep{Drakos2022}  are a $1 \deg^2$ lightcone containing galaxies with stellar masses $M>10^5$\,M$_\odot$ beyond redshift $z=10$. These catalogs are based on a high-resolution dark matter-only simulation, and populated with galaxies using subhalo abundance matching (SHAM). Galaxy stellar and morphological parameters were assigned using observed and theoretical scaling relations (similar to the semi-empirical used in JAGUAR \citep[JAdes extraGalactic Ultradeep Artificial Realizations,][as described in Section~\ref{sec:JAGUAR}]{Williams2018}. Spectra were generated using the Flexible Stellar Popular Synthesis (FSPS) code \citep{Conroy2009b,Conroy2010}, assuming a \cite{ChabrierIMF} initial mass function. The spectra modelling includes IGM absorption \citep{Madau1995}, nebular emission \citep{Byler2017}, dust emission \citep{Draine2007}, dust absorption \citep{Calzetti2000} and asymptotic giant branch circumstellar dust \citep{Villaume2015}.

\subsubsection{Santa Cruz semi-analytical model}

The Santa Cruz semi-analytical model \citep[SCSAM,][]{Somerville2015, Yung2019} is a physically, computationally efficient modelling framework that tracks the formation and evolution of galaxies in dark matter halo merger trees under the influence of a set of carefully curated physical processes, including cosmological accretion, cooling, star formation, chemical enrichment, and stellar and AGN feedback. The model configuration and physical parameters are based on the calibration from \citet{Yung2019} and \citet{Yung2021}. The models have been shown to reproduce the observed evolution in high-redshift (e.g. $z \gtrsim 4$) one-point distribution functions of $M_\text{UV}$, $M_*$, and SFR \citep{Yung2019, Yung2019b}, cosmic reionization constraints \citep{Yung2020a, Yung2020b}, as well as two-point auto-correlation functions from $0 < z < 7.5$ \citep{Yung2022, Yung2023}. This work adopted an augmented version of the mock galaxy catalogues presented in \citet[][also see \citealt{Somerville2021}]{Yung2022}, which spans $0 < z < 10$ over a total area of 782 arcmin$^2$ with coordinates overlapping with the observed EGS / CEERS field. This lightcone contains galaxies in rest-frame $M_\text{UV}$ range between $-16 \gtrsim M_\text{UV} \gtrsim -22$. We refer the reader to \citet{Yung2022} for an overview for the internal workflow of the Santa Cruz SAM and the specification of the lightcone.

Synthetic SEDs for stellar population are constructed for individual galaxies based on their predicted star formation and chemical enrichment histories and on stellar the population synthesis models of \citet{BC03}. In addition, nebular emission is modelled using the approach described in \citet{Hirschmann2017, Hirschmann2019, Hirschmann2022}, which self-consistently predicted nebular emission lines excited by young stars, AGN, and post-AGB stellar populations in the high-resolution synthetic spectra and the broad- and medium-band photometry (Yung, Hirschmann, Somerville et al. \textit{in preparation}).

\subsubsection{FLARES}

FLARES \citep[the First Light And Reionisation Epoch Simulations][]{FLARES-I} is a suite of high-redshift $z>5$ hydrodynamical re-simulations based on the EAGLE physics model \citep{schaye_eagle_2015, crain_eagle_2015}. The strategy adopted by FLARES was chosen to enable the study of a wide range of environments and extend the dynamic range of EAGLE predictions. For example, FLARES simulates around 100$\times$ as many massive ($M_{\star}>10^{10}\ {\rm M_{\odot}}$) galaxies compared to the EAGLE reference simulation. The synthetic photometry is described in \citet{FLARES-II} and predictions for the colour evolution of galaxies is presented in \citet{FLARES-VI}. In short, each star particle is associated with a spectral energy distribution based on its age and metallicity assuming a \citet{ChabrierIMF} IMF and v2.2.1 of the Binary Population And Spectral Synthesis \citep[BPASS,][]{BPASS2.2.1} SPS code. Each star particle is assumed to photoionise a region surrounding it giving rise to nebular emission which is modelled using version 17.03 of the \textsc{cloudy} photoionisation model \citep{Cloudy17.02}. Dust attenuation is modelled on a particle-by-particle basis by calculating the line-of-sight surface density of metals, which is assumed to track dust, and converting this to an optical depth. The model also applies additional dust attenuation to young stars (age$\le10$Myr), under the assumption that they are still embedded in their nascent birth clouds \cite[]{Charlot_and_Fall}.

As FLARES utilised a series of integer redshift snapshots, colours at intermediate redshifts were calculated by using the rest-frame SEDs of galaxies at the nearest snapshot but using the specific redshift to calculate observed-frame SEDs \citep[see,][]{FLARES-VI}.


\subsection{Sample definition}

The four models described above all provide synthetic model photometry in the \emph{HST} and \emph{JWST} bands observed by CEERS. To align these model predictions with our observed sample we create a new sample which is constructed using the same methodology as applied to the observations.

To do this, we first apply photometric noise to the simulated galaxies. To do this we measure the average noise as a function of flux for the entire CEERS sample in each band and fit the resulting relation by a power-law. We then use this power-law to determine a unique value of the standard deviation of the noise for each object in each band and use this to add random noise. As discussed in \S\ref{sec:conclusions:future} a better approach would be to insert synthetic images of the simulated galaxies into the original images and run the full observational pipeline. We defer this to a future dedicated study which also incorporates a comparison of the observed and predicted morphology.

\begin{figure}
    \includegraphics[width=\columnwidth]{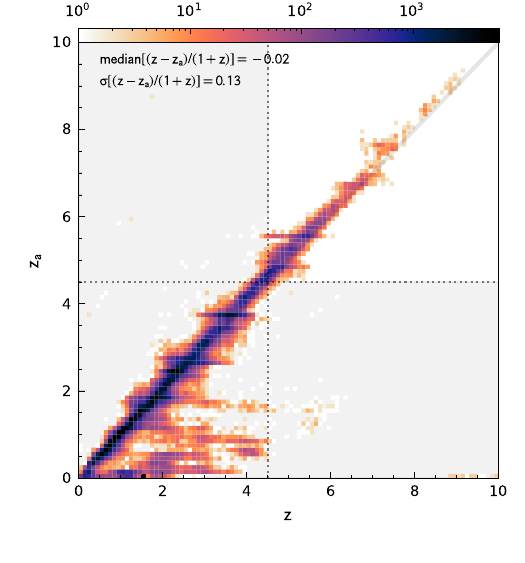}
    \includegraphics[width=\columnwidth]{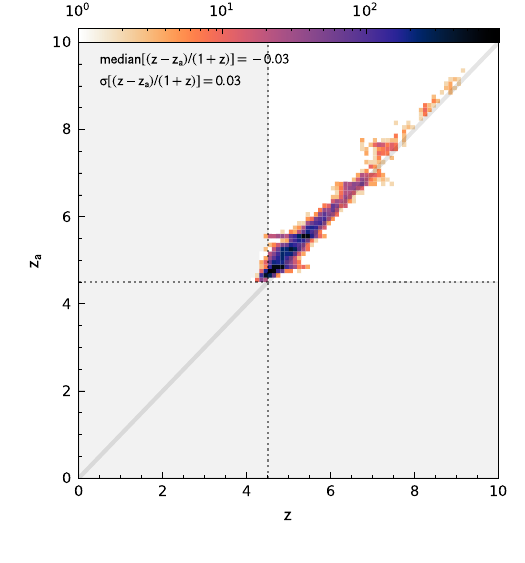}
	\caption{A density plot (with density colour coded on a log scale) showing the relationship between the true and photometric redshift for the Santa Cruz SAM based sample. The top-panel shows all objects with F277W$>50$ nJy and the bottom panel shows just objects meeting our selection criteria. 
	\label{fig:z_pz}}
\end{figure}

\begin{figure}
 	\includegraphics[width=\columnwidth]{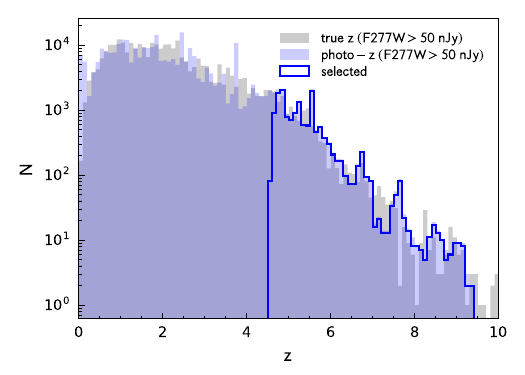}
	\caption{The distribution of true and photometric redshifts of galaxies meeting our flux criteria in Santa Cruz SAM.
	\label{fig:z_dist}}
\end{figure}

Next, we pass this \emph{noisy} photometry to \textsc{eazy} providing $P(z)$ for each simulated source, thus allowing us to apply the same selection criteria applied to the observations (see \S\ref{sec:obs:pz}). In Figure \ref{fig:z_pz} we show the relationship between the true redshift and the best photometric redshift estimate ($z_a$), utilising the Santa Cruz SAM synthetic observations, for both the full flux-limited sample and the sample with our additional cuts. The correspondence for the full sample is reasonably good with bias $-0.02$ and scatter $0.13$. For our selected high-redshift sample the scatter is $0.03$. As can be seen in Figure \ref{fig:z_pz} the effect of applying our selection eliminates low-redshift contamination (sources in the upper-left quadrant) except sources near the boundary. These figures also reveal a handful of true $z>4.5$ galaxies which have photometric redshifts at $z<4.5$, and thus not subsequently selected (lower-right quadrant). However, at least for those at $z<8$, these galaxies make up only a small fraction of the total sample. In fact, this does not actually reduce the number of selected galaxies since some galaxies scatter in from fainter fluxes and lower redshifts since there are more objects at immediately lower redshift and flux. This can be seen in Figure \ref{fig:z_dist} where we show the true redshift distribution, the distribution of photometric redshifts, and the distribution of photometric redshifts for our selected sample. While the impact at $z<8$ is negligible we do appear to lose a large fraction of our extreme redshift sources ($z>9$). It is important to note that these results are not unique to the Santa Cruz SAM; the other models considered exhibit similar trends.

Armed with \emph{noisy} model photometry and photometric redshifts we can measure the flux distribution and colour evolution in the same way as the observations. These are presented for the four models alongside the observational measurements in Section \ref{sec:results}.

\section{Results}\label{sec:results}

\subsection{Flux distribution}\label{sec:results:magnitude}

\begin{figure*}
 	\includegraphics[width=2\columnwidth]{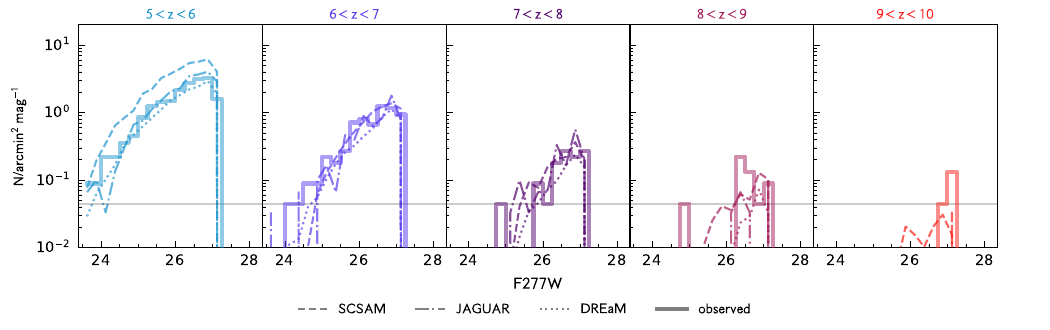}
    \includegraphics[width=2\columnwidth]{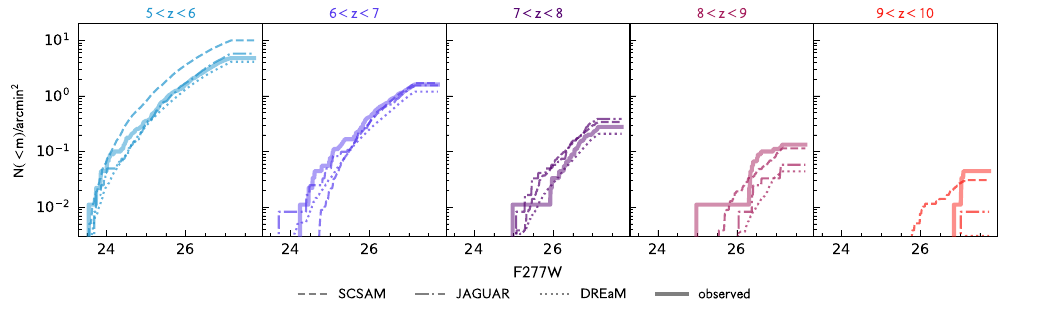}
	\caption{The F277W differential (top) and cumulative (bottom) magnitude distribution of both the observed (solid thick faint line) and model (various thin dark lines) samples. The horizontal line on the differential plot denotes the number counts expected for the observation of a single object in the bin.
	\label{fig:mz}}
\end{figure*}

We begin, in Figure \ref{fig:mz}, by plotting the F277W magnitude distribution of both the observed sample and model samples using the Santa Cruz SAM, JAGUAR, and DREaM lightcones\footnote{Since the FLARES simulation does not provide a lightcone here we do not compare against predictions from that model.}. As anticipated from earlier work the observed distribution increases rapidly to lower flux before sharply dropping at F277W$>27$. The sharpness of our drop here simply reflects our selection criteria, specifically our flux limit. 

At $6<z<8$ this reveals good agreement between the observations and all three models. At $5<z<6$ both DREaM and JAGUAR produce a good match to the observed flux distribution while the Santa Cruz SAM predicts somewhat more galaxies. At $z>8$ the observed sample differs somewhat from the model distributions, however, this may simply reflect the small numbers of galaxies in our selection at these redshifts.

\subsection{Colours}\label{sec:results:colours}

\begin{figure}
 	\includegraphics[width=\columnwidth]{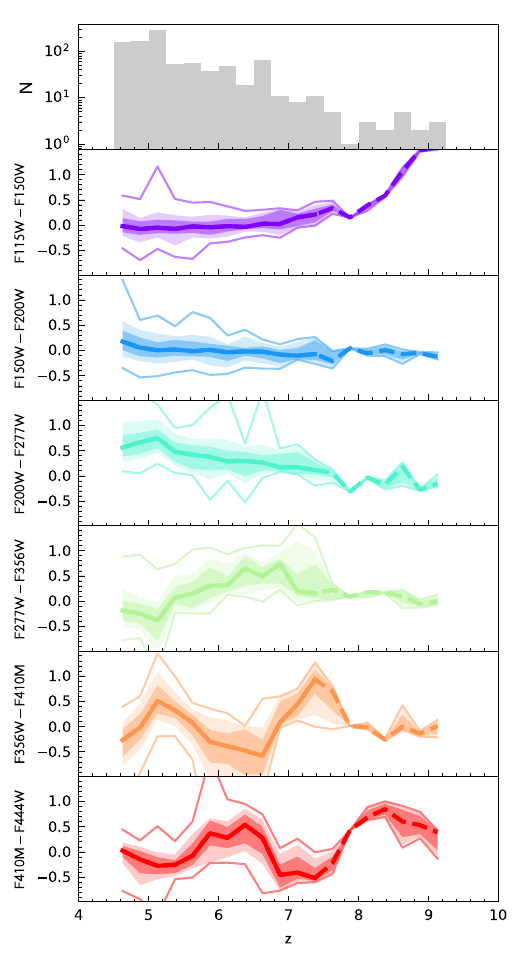}
	\caption{The observed colour evolution of our full sample in the six consecutive NIRCam colours. Here we plot the statistics of the colour distribution in each $\Delta z=0.25$ redshift bin. The thick solid line shows the median colour in bins containing $>5$ galaxies while the dashed line shows the median for bins with $N=[1,5]$. The two shaded regions show the 68\% and 95\% ranges while the two thin lines show the minimum and maximum values. The top-panel shows the redshift distribution of sources on a logarithmic scale. Figure \ref{fig:cz_models} also shows the error on the median. 
	\label{fig:cz_obs}}
\end{figure}

We next explore the redshift evolution of the observed and predicted NIRCam colours of galaxies. Here we begin, in Figure \ref{fig:cz_obs}, by showing the observed colour distribution for six consecutive NIRCam colours available to CEERS. We show the median, central 68\% and 95\% ranges, and min/max values. 

This figure immediately reveals the signature of strong nebular line emission in the F356W$-$F410M and F410M$-$F444W, and the presence of the Lyman-$\alpha$ break, as expected from the modelling presented in Section \ref{sec:theory}. 
There is some hint that the colour distribution of galaxies, at fixed redshift, deviates from a simple Gaussian, with more sources having extreme values relative to the median than expected. Indeed, there are also a handful of sources that have colours that deviate significantly from the median relative to the central 68\% range. On inspection of these sources, there is nothing to suggest from their morphology or SEDs that they are not compelling candidates, however, without spectroscopic confirmation we cannot rule out the possibility that they are hitherto unidentified contaminants.

\subsubsection{Luminosity dependence}\label{sec:predictions:colours:luminosity}

\begin{figure}
 	\includegraphics[width=\columnwidth]{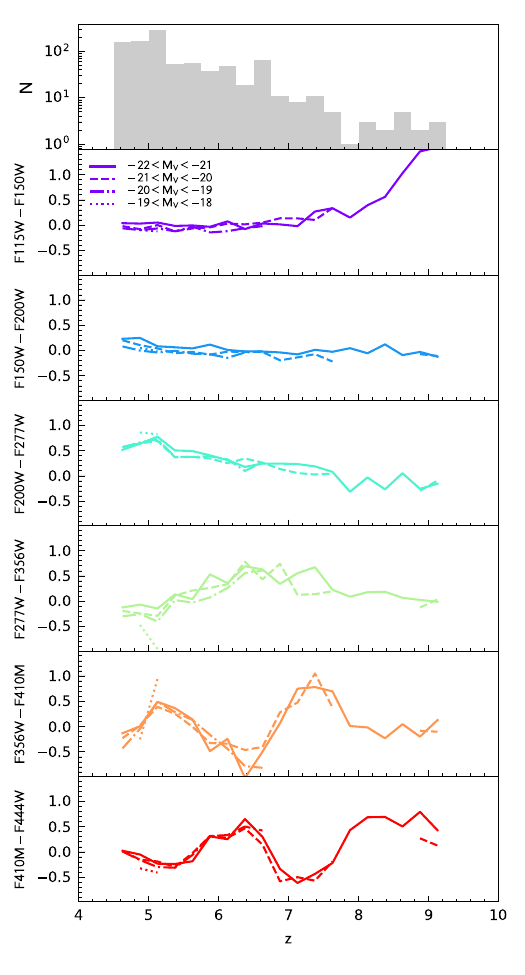}
	\caption{The colour evolution of our observed CEERS sample split by the rest-frame V-band absolute magnitude.
	\label{fig:cz_obs_M}}
\end{figure}

Next, in Figure \ref{fig:cz_obs_M} we again show the median colour evolution of our observed sample but this time we split the sample into rest-frame V-band luminosity (absolute magnitude) bins over $M_{V}=[-18,-22]$. The median colour in each luminosity bin is statistically consistent with the average of the whole sample suggesting little evidence of strong luminosity dependence. However, this is a relatively small sample covering a limited luminosity range thus it is difficult to draw strong conclusions.

\subsubsection{Comparison with model predictions}\label{sec:predictions:colours:models}

\begin{figure}
 	\includegraphics[width=\columnwidth]{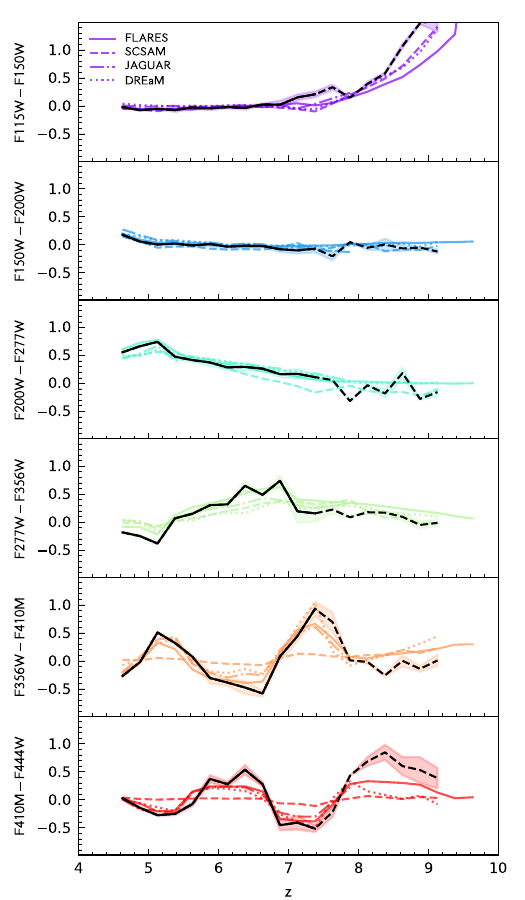}
	\caption{The colour evolution predicted by the four models described in \S\ref{sec:models.models} alongside the observational measurement from CEERS. For the models, we show only the median colour and only for bins with $>5$ objects. For the observations we, again, show the median colour (in black) but show both when $N>5$ (solid line) and $N=[1,5]$ (dashed line). For the observational sample, we also show an estimate of the standard error on the median, calculated as $1.25\sigma/\sqrt{N}$.
	\label{fig:cz_models}}
\end{figure}

We now, in Figure \ref{fig:cz_models}, turn our attention to the comparison of our observed average (median) colour evolution to the four models described in \S\ref{sec:models.models}. Broadly all the models provide relatively good agreement with the observations, particularly the colours not impacted by strong line emissison (F115W$-$F150W, F150W$-$F200W, F200W$-$F277W, F277W$-$F356W). However, there are some differences and discrepancies with the observed colour evolution.

First, there is a difference in the magnitude of the emission line-driven features in the F356W$-$F410M and F410M$-$F444W colours. As noted in Section \ref{sec:theory} these colours trace the equivalent width of the strong H$\alpha$ ($z\sim 5-6$) or \OIIIHb\ ($z=7-9$) lines. Here FLARES, JAGUAR, and DREaM perform relatively well, capturing the shape and normalisation of the average evolution within the statistical uncertainty, at least at $z<8$. The outlier here is the Santa Cruz SAM which does not exhibit the redshift evolution indicative of strong line emission. The cause of this remains unclear and is currently been investigated.

While FLARES, JAGUAR, and DREaM produce a good match to the observations at $z<8$ no model matches the magnitude of the shift in the F410M$-$F444W colour at $z>8$. If real, this suggests the observed galaxies in this regime have stronger \OIIIHb\ line emission than predicted by any model. This could be a consequence of a larger ionisation parameter $U$ than assumed or a more fundamental issue with the star formation physics such as a shift to a high-mass biased initial mass function, evidence of which has recently been presented by \citet{Cameron23} albeit at lower-redshift. However, there are relatively few sources in our sample at these redshifts, meaning we cannot yet rule out a statistical fluctuation. 

Finally, another difference is the predicted strength of the Lyman-break, probed by the F115W$-$F150W colour at $z>8$. Here all the models, but particularly FLARES, under-predict the colour. This possibly suggests that the IGM attenuation or Lyman-$\alpha$ emission is not appropriately modelled. 

\section{Conclusions}\label{sec:conclusions}

In this work, we have assembled a sample of observed galaxies at $5<z<10$ using the Cosmic Evolution Early Release Science (CEERS) survey and measured the redshift evolution and luminosity dependence of their colours. In parallel we have applied the same selection procedure, including adding appropriate noise and constraining the photometric redshift, to synthetic observations from the DREAM and JAGUAR semi-empirical models, the Santa Cruz semi-analytical model, and the hydrodynamical First Light And Reionisation Epoch Simulations (FLARES). Our conclusions are:

\begin{itemize}

    \item There is good agreement, at $5<z<8$, between the observed and model F277W magnitude distribution. At $z>8$ the agreement weakens but this may reflect the small sample size currently available at the higher-redshifts.

    \item The observed statistical colour evolution shows clear evidence for the presence of strong line emission.
        
    \item The observed colour evolution shows little dependence on luminosity, at least over the ranges probed by our sample.

    \item The closest agreement with the observations comes from the FLARES hydrodynamical simulations and the JAGUAR semi-empirical model. Both produce a good match to the observed colour evolution at $z<8$. However, all the models considered under-predict the F410M$-$F444W colour at $z>8$, suggesting a stronger contribution of \OIIIHb\ line emission than predicted. This could be due to a change in the properties of the H\textsc{ii} regions (e.g. an increased ionisation parameter) or something more fundamental such as a shift to a high-mass biased (top-heavy) initial mass function. However, in this analysis, the number of galaxies at $z>8$ is small meaning we can't yet rule out a statistical fluctuation. 
    
\end{itemize}

\subsection{Future work}\label{sec:conclusions:future}

The work presented here represents an initial exploration of the spectral energy distribution of galaxies at high-redshift and their comparison with galaxy formation models. In future work (Turner et al. \emph{in-prep}) we will explore a number of refinements:

\begin{itemize}

\item Firstly, \emph{JWST} has only just embarked on its potentially 20-year observing programme during which it will not only observe substantially deeper but also wider, dramatically expanding the number and dynamic range of high-redshift galaxy samples. Indeed, at the time of writing several other programmes are already completed or underway. A future iteration of this analysis will expand this methodology to encompass other available datasets. 

\item Second, in this work we have adopted a single methodology to identify sources, measure their photometry, and constrain their photometric redshifts. Since multiple approaches exist in the literature we will, in this future work, explore alternatives to understand the impact of these assumptions. This includes the source extraction tool (and parameters), the choice of method for measuring colours, and the photometric redshift code.

\item Third, in this work we use catalogues of synthetic sources adding noise based on a fit to the observations. A better approach is to insert the synthetic sources into the actual observations and subsequently treat them as real sources. This is significantly more challenging both because it requires that we construct not only synthetic photometry but synthetic images but also that it will require multiple runs of the source extraction and photometry pipelines over the real images, incurring significant computational expense. This approach would also enable direct comparisons of predicted and observed morphologies providing an additional constraint on the model physics \citep[e.g. see][]{FLARES-IX}. 

\item Fourth, the models considered in this work, in addition to adopting fundamentally different modelling approaches (semi-empirical vs. semi-analytical vs. fully hydrodynamical), also apply heterogeneous approaches to modelling the spectral energy distributions of galaxies. This includes assuming different stellar population synthesis models and initial mass functions. A better approach would be forward-model the different models using a consistent approach while also exploring the impact of these assumptions within the realms of that allowed by the model. This is a core aim of the \synthesizer\footnote{\url{https://github.com/flaresimulations/synthesizer}} package and project.

\item Finally, in this work we have compared individual colours with models, ignoring cross-correlations. However, using some form of dimensionality reduction (e.g. principle component analysis or uniform manifold approximation and projection), may enable simultaneously comparing a number of colours (and other information) providing a more complete comparison with observations 

\end{itemize}
	
\section*{Acknowledgements}




SMW and WJR thank STFC for support through ST/X001040/1. JT, JD, and LTCS is supported by an STFC studentship. DI acknowledges support by the European Research Council via ERC Consolidator Grant KETJU (no. 818930) and the CSC – IT Center for Science, Finland. The Cosmic Dawn Center (DAWN) is funded by the Danish National Research Foundation under grant DNRF140.

We also wish to acknowledge the following open source software packages used in the analysis: \textsc{Scipy} \cite[][]{2020SciPy-NMeth}, \textsc{Astropy} \cite[][]{robitaille_astropy:_2013}, and \textsc{Matplotlib} \cite[][]{Hunter:2007}. 

We list here the roles and contributions of the authors according to the Contributor Roles Taxonomy (CRediT)\footnote{\url{https://credit.niso.org/}}.
\textbf{SMW, JT}: Conceptualization, Data curation, Methodology, Investigation, Formal Analysis, Visualization, Writing - original draft.
\textbf{MB, SF}: Data curation, Writing - review \& editing.
\textbf{RA, PB, RB, AD, JD, NED, FF, NPH, BH, DI, AMK, CL, EM, WJR, LTCS, APV, LYAY}: Writing - review \& editing.

\section*{Data Availability Statement}



\bibliographystyle{mnras}
\bibliography{ceers-colours, flares} 

\begin{thebibliography}{}
\makeatletter
\relax
\def\mn@urlcharsother{\let\do\@makeother \do\$\do\&\do\#\do\^\do\_\do\%\do\~}
\def\mn@doi{\begingroup\mn@urlcharsother \@ifnextchar [ {\mn@doi@}
  {\mn@doi@[]}}
\def\mn@doi@[#1]#2{\def\@tempa{#1}\ifx\@tempa\@empty \href
  {http://dx.doi.org/#2} {doi:#2}\else \href {http://dx.doi.org/#2} {#1}\fi
  \endgroup}
\def\mn@eprint#1#2{\mn@eprint@#1:#2::\@nil}
\def\mn@eprint@arXiv#1{\href {http://arxiv.org/abs/#1} {{\tt arXiv:#1}}}
\def\mn@eprint@dblp#1{\href {http://dblp.uni-trier.de/rec/bibtex/#1.xml}
  {dblp:#1}}
\def\mn@eprint@#1:#2:#3:#4\@nil{\def\@tempa {#1}\def\@tempb {#2}\def\@tempc
  {#3}\ifx \@tempc \@empty \let \@tempc \@tempb \let \@tempb \@tempa \fi \ifx
  \@tempb \@empty \def\@tempb {arXiv}\fi \@ifundefined
  {mn@eprint@\@tempb}{\@tempb:\@tempc}{\expandafter \expandafter \csname
  mn@eprint@\@tempb\endcsname \expandafter{\@tempc}}}

\bibitem[\protect\citeauthoryear{{Bagley} et~al.,}{{Bagley}
  et~al.}{2023}]{Bagley2023}
{Bagley} M.~B.,  et~al., 2023, \mn@doi [\apjl] {10.3847/2041-8213/acbb08},
  \href {https://ui.adsabs.harvard.edu/abs/2023ApJ...946L..12B} {946, L12}

\bibitem[\protect\citeauthoryear{{Bertin} \& {Arnouts}}{{Bertin} \&
  {Arnouts}}{1996}]{Bertin1996}
{Bertin} E.,  {Arnouts} S.,  1996, \mn@doi [\aaps] {10.1051/aas:1996164}, \href
  {https://ui.adsabs.harvard.edu/abs/1996A&AS..117..393B} {117, 393}

\bibitem[\protect\citeauthoryear{{Blaizot}, {Wadadekar}, {Guiderdoni},
  {Colombi}, {Bertin}, {Bouchet}, {Devriendt}  \& {Hatton}}{{Blaizot}
  et~al.}{2005}]{Blaizot2005}
{Blaizot} J.,  {Wadadekar} Y.,  {Guiderdoni} B.,  {Colombi} S.~T.,  {Bertin}
  E.,  {Bouchet} F.~R.,  {Devriendt} J. E.~G.,   {Hatton} S.,  2005, \mn@doi
  [\mnras] {10.1111/j.1365-2966.2005.09019.x}, \href
  {https://ui.adsabs.harvard.edu/abs/2005MNRAS.360..159B} {360, 159}

\bibitem[\protect\citeauthoryear{Brammer, van Dokkum  \& Coppi}{Brammer
  et~al.}{2008}]{Brammer2008}
Brammer G.~B.,  van Dokkum P.~G.,   Coppi P.,  2008, The Astrophysical Journal,
  686, 1503

\bibitem[\protect\citeauthoryear{{Bravo}, {Lagos}, {Robotham}, {Bellstedt}  \&
  {Obreschkow}}{{Bravo} et~al.}{2020}]{Bravo2020}
{Bravo} M.,  {Lagos} C. d.~P.,  {Robotham} A. S.~G.,  {Bellstedt} S.,
  {Obreschkow} D.,  2020, \mn@doi [\mnras] {10.1093/mnras/staa2027}, \href
  {https://ui.adsabs.harvard.edu/abs/2020MNRAS.497.3026B} {497, 3026}

\bibitem[\protect\citeauthoryear{{Bruzual} \& {Charlot}}{{Bruzual} \&
  {Charlot}}{2003}]{BC03}
{Bruzual} G.,  {Charlot} S.,  2003, \mn@doi [\mnras]
  {10.1046/j.1365-8711.2003.06897.x}, \href
  {https://ui.adsabs.harvard.edu/abs/2003MNRAS.344.1000B} {344, 1000}

\bibitem[\protect\citeauthoryear{{Bushouse} et~al.,}{{Bushouse}
  et~al.}{2022}]{jwstpipeline}
{Bushouse} H.,  et~al., 2022, {JWST Calibration Pipeline}, Zenodo,
  \mn@doi{10.5281/zenodo.7325378}

\bibitem[\protect\citeauthoryear{{Byler}, {Dalcanton}, {Conroy}  \&
  {Johnson}}{{Byler} et~al.}{2017}]{Byler2017}
{Byler} N.,  {Dalcanton} J.~J.,  {Conroy} C.,   {Johnson} B.~D.,  2017, \mn@doi
  [\apj] {10.3847/1538-4357/aa6c66}, \href
  {https://ui.adsabs.harvard.edu/abs/2017ApJ...840...44B} {840, 44}

\bibitem[\protect\citeauthoryear{{Calzetti}, {Armus}, {Bohlin}, {Kinney},
  {Koornneef}  \& {Storchi-Bergmann}}{{Calzetti} et~al.}{2000}]{Calzetti2000}
{Calzetti} D.,  {Armus} L.,  {Bohlin} R.~C.,  {Kinney} A.~L.,  {Koornneef} J.,
   {Storchi-Bergmann} T.,  2000, \mn@doi [\apj] {10.1086/308692}, \href
  {https://ui.adsabs.harvard.edu/abs/2000ApJ...533..682C} {533, 682}

\bibitem[\protect\citeauthoryear{{Cameron}, {Katz}, {Witten}, {Saxena},
  {Laporte}  \& {Bunker}}{{Cameron} et~al.}{2023}]{Cameron23}
{Cameron} A.~J.,  {Katz} H.,  {Witten} C.,  {Saxena} A.,  {Laporte} N.,
  {Bunker} A.~J.,  2023, \mn@doi [arXiv e-prints] {10.48550/arXiv.2311.02051},
  \href {https://ui.adsabs.harvard.edu/abs/2023arXiv231102051C} {p.
  arXiv:2311.02051}

\bibitem[\protect\citeauthoryear{{Carnall}, {Leja}, {Johnson}, {McLure},
  {Dunlop}  \& {Conroy}}{{Carnall} et~al.}{2019}]{Carnall2019}
{Carnall} A.~C.,  {Leja} J.,  {Johnson} B.~D.,  {McLure} R.~J.,  {Dunlop}
  J.~S.,   {Conroy} C.,  2019, \mn@doi [\apj] {10.3847/1538-4357/ab04a2}, \href
  {https://ui.adsabs.harvard.edu/abs/2019ApJ...873...44C} {873, 44}

\bibitem[\protect\citeauthoryear{{Carrasco}, {Trenti}, {Mutch}  \&
  {Oesch}}{{Carrasco} et~al.}{2018}]{GLACiAR}
{Carrasco} D.,  {Trenti} M.,  {Mutch} S.,   {Oesch} P.~A.,  2018, \mn@doi
  [\pasa] {10.1017/pasa.2018.17}, \href
  {https://ui.adsabs.harvard.edu/abs/2018PASA...35...22C} {35, e022}

\bibitem[\protect\citeauthoryear{Chabrier}{Chabrier}{2003a}]{chabrier_galactic_2003}
Chabrier G.,  2003a, \mn@doi [\pasp] {10.1086/376392}, 115, 763

\bibitem[\protect\citeauthoryear{Chabrier}{Chabrier}{2003b}]{ChabrierIMF}
Chabrier G.,  2003b, \mn@doi [\pasp] {10.1086/376392}, 115, 763

\bibitem[\protect\citeauthoryear{{Charlot} \& {Fall}}{{Charlot} \&
  {Fall}}{2000}]{Charlot_and_Fall}
{Charlot} S.,  {Fall} S.~M.,  2000, \mn@doi [\apj] {10.1086/309250}, \href
  {https://ui.adsabs.harvard.edu/abs/2000ApJ...539..718C} {539, 718}

\bibitem[\protect\citeauthoryear{{Chevallard} \& {Charlot}}{{Chevallard} \&
  {Charlot}}{2016}]{Chevallard2016}
{Chevallard} J.,  {Charlot} S.,  2016, \mn@doi [\mnras]
  {10.1093/mnras/stw1756}, \href
  {https://ui.adsabs.harvard.edu/abs/2016MNRAS.462.1415C} {462, 1415}

\bibitem[\protect\citeauthoryear{{Conroy}}{{Conroy}}{2013}]{Conroy2013}
{Conroy} C.,  2013, \mn@doi [\araa] {10.1146/annurev-astro-082812-141017},
  \href {https://ui.adsabs.harvard.edu/abs/2013ARA&A..51..393C} {51, 393}

\bibitem[\protect\citeauthoryear{{Conroy} \& {Gunn}}{{Conroy} \&
  {Gunn}}{2010}]{Conroy2010}
{Conroy} C.,  {Gunn} J.~E.,  2010, {FSPS: Flexible Stellar Population
  Synthesis}, Astrophysics Source Code Library, record ascl:1010.043
  (\mn@eprint {ascl} {1010.043})

\bibitem[\protect\citeauthoryear{{Conroy}, {Gunn}  \& {White}}{{Conroy}
  et~al.}{2009a}]{Conroy2009}
{Conroy} C.,  {Gunn} J.~E.,   {White} M.,  2009a, \mn@doi [\apj]
  {10.1088/0004-637X/699/1/486}, \href
  {https://ui.adsabs.harvard.edu/abs/2009ApJ...699..486C} {699, 486}

\bibitem[\protect\citeauthoryear{{Conroy}, {Gunn}  \& {White}}{{Conroy}
  et~al.}{2009b}]{Conroy2009b}
{Conroy} C.,  {Gunn} J.~E.,   {White} M.,  2009b, \mn@doi [\apj]
  {10.1088/0004-637X/699/1/486}, \href
  {https://ui.adsabs.harvard.edu/abs/2009ApJ...699..486C} {699, 486}

\bibitem[\protect\citeauthoryear{Crain et~al.,}{Crain
  et~al.}{2015}]{crain_eagle_2015}
Crain R.~A.,  et~al., 2015, \mn@doi [\mnras] {10.1093/mnras/stv725}, 450, 1937

\bibitem[\protect\citeauthoryear{{Draine} \& {Li}}{{Draine} \&
  {Li}}{2007}]{Draine2007}
{Draine} B.~T.,  {Li} A.,  2007, \mn@doi [\apj] {10.1086/511055}, \href
  {https://ui.adsabs.harvard.edu/abs/2007ApJ...657..810D} {657, 810}

\bibitem[\protect\citeauthoryear{{Drakos} et~al.,}{{Drakos}
  et~al.}{2022}]{Drakos2022}
{Drakos} N.~E.,  et~al., 2022, \mn@doi [\apj] {10.3847/1538-4357/ac46fb}, \href
  {https://ui.adsabs.harvard.edu/abs/2022ApJ...926..194D} {926, 194}

\bibitem[\protect\citeauthoryear{{Endsley}, {Stark}, {Whitler}, {Topping},
  {Chen}, {Plat}, {Chisholm}  \& {Charlot}}{{Endsley} et~al.}{2023}]{Endsley23}
{Endsley} R.,  {Stark} D.~P.,  {Whitler} L.,  {Topping} M.~W.,  {Chen} Z.,
  {Plat} A.,  {Chisholm} J.,   {Charlot} S.,  2023, \mn@doi [\mnras]
  {10.1093/mnras/stad1919}, \href
  {https://ui.adsabs.harvard.edu/abs/2023MNRAS.524.2312E} {524, 2312}

\bibitem[\protect\citeauthoryear{{Ferland} et~al.,}{{Ferland}
  et~al.}{2013}]{cloudy13}
{Ferland} G.~J.,  et~al., 2013, \mn@doi [\rmxaa] {10.48550/arXiv.1302.4485},
  \href {https://ui.adsabs.harvard.edu/abs/2013RMxAA..49..137F} {49, 137}

\bibitem[\protect\citeauthoryear{{Ferland} et~al.,}{{Ferland}
  et~al.}{2017}]{Cloudy17.02}
{Ferland} G.~J.,  et~al., 2017, \rmxaa, \href
  {https://ui.adsabs.harvard.edu/abs/2017RMxAA..53..385F} {53, 385}

\bibitem[\protect\citeauthoryear{{Finkelstein} et~al.,}{{Finkelstein}
  et~al.}{2023}]{Finkelstein2023}
{Finkelstein} S.~L.,  et~al., 2023, \mn@doi [\apjl] {10.3847/2041-8213/acade4},
  \href {https://ui.adsabs.harvard.edu/abs/2023ApJ...946L..13F} {946, L13}

\bibitem[\protect\citeauthoryear{{Fortuni} et~al.,}{{Fortuni}
  et~al.}{2023}]{Fortuni23}
{Fortuni} F.,  et~al., 2023, \mn@doi [arXiv e-prints]
  {10.48550/arXiv.2305.19166}, \href
  {https://ui.adsabs.harvard.edu/abs/2023arXiv230519166F} {p. arXiv:2305.19166}

\bibitem[\protect\citeauthoryear{{Grogin} et~al.,}{{Grogin}
  et~al.}{2011}]{Grogin2011}
{Grogin} N.~A.,  et~al., 2011, \mn@doi [\apjs] {10.1088/0067-0049/197/2/35},
  \href {https://ui.adsabs.harvard.edu/abs/2011ApJS..197...35G} {197, 35}

\bibitem[\protect\citeauthoryear{{Hirschmann}, {Charlot}, {Feltre}, {Naab},
  {Choi}, {Ostriker}  \& {Somerville}}{{Hirschmann}
  et~al.}{2017}]{Hirschmann2017}
{Hirschmann} M.,  {Charlot} S.,  {Feltre} A.,  {Naab} T.,  {Choi} E.,
  {Ostriker} J.~P.,   {Somerville} R.~S.,  2017, \mn@doi [\mnras]
  {10.1093/mnras/stx2180}, \href
  {https://ui.adsabs.harvard.edu/abs/2017MNRAS.472.2468H} {472, 2468}

\bibitem[\protect\citeauthoryear{{Hirschmann}, {Charlot}, {Feltre}, {Naab},
  {Somerville}  \& {Choi}}{{Hirschmann} et~al.}{2019}]{Hirschmann2019}
{Hirschmann} M.,  {Charlot} S.,  {Feltre} A.,  {Naab} T.,  {Somerville} R.~S.,
   {Choi} E.,  2019, \mn@doi [\mnras] {10.1093/mnras/stz1256}, \href
  {https://ui.adsabs.harvard.edu/abs/2019MNRAS.487..333H} {487, 333}

\bibitem[\protect\citeauthoryear{{Hirschmann} et~al.,}{{Hirschmann}
  et~al.}{2022}]{Hirschmann2022}
{Hirschmann} M.,  et~al., 2022, \mn@doi [arXiv e-prints]
  {10.48550/arXiv.2212.02522}, \href
  {https://ui.adsabs.harvard.edu/abs/2022arXiv221202522H} {p. arXiv:2212.02522}

\bibitem[\protect\citeauthoryear{Hunter}{Hunter}{2007}]{Hunter:2007}
Hunter J.~D.,  2007, \mn@doi [Computing in Science \& Engineering]
  {10.1109/MCSE.2007.55}, 9, 90

\bibitem[\protect\citeauthoryear{{Koekemoer} et~al.,}{{Koekemoer}
  et~al.}{2011}]{Koekemoer2011}
{Koekemoer} A.~M.,  et~al., 2011, \mn@doi [\apjs] {10.1088/0067-0049/197/2/36},
  \href {https://ui.adsabs.harvard.edu/abs/2011ApJS..197...36K} {197, 36}

\bibitem[\protect\citeauthoryear{{Laigle} et~al.,}{{Laigle}
  et~al.}{2019}]{Laigle2019}
{Laigle} C.,  et~al., 2019, \mn@doi [\mnras] {10.1093/mnras/stz1054}, \href
  {https://ui.adsabs.harvard.edu/abs/2019MNRAS.486.5104L} {486, 5104}

\bibitem[\protect\citeauthoryear{{Larson} et~al.,}{{Larson}
  et~al.}{2022}]{Larson2022}
{Larson} R.~L.,  et~al., 2022, \mn@doi [arXiv e-prints]
  {10.48550/arXiv.2211.10035}, \href
  {https://ui.adsabs.harvard.edu/abs/2022arXiv221110035L} {p. arXiv:2211.10035}

\bibitem[\protect\citeauthoryear{{Lovell}, {Vijayan}, {Thomas}, {Wilkins},
  {Barnes}, {Irodotou}  \& {Roper}}{{Lovell} et~al.}{2021}]{FLARES-I}
{Lovell} C.~C.,  {Vijayan} A.~P.,  {Thomas} P.~A.,  {Wilkins} S.~M.,  {Barnes}
  D.~J.,  {Irodotou} D.,   {Roper} W.,  2021, \mn@doi [\mnras]
  {10.1093/mnras/staa3360}, \href
  {https://ui.adsabs.harvard.edu/abs/2021MNRAS.500.2127L} {500, 2127}

\bibitem[\protect\citeauthoryear{{Lower}, {Narayanan}, {Leja}, {Johnson},
  {Conroy}  \& {Dav{\'e}}}{{Lower} et~al.}{2020}]{Lower2020}
{Lower} S.,  {Narayanan} D.,  {Leja} J.,  {Johnson} B.~D.,  {Conroy} C.,
  {Dav{\'e}} R.,  2020, \mn@doi [\apj] {10.3847/1538-4357/abbfa7}, \href
  {https://ui.adsabs.harvard.edu/abs/2020ApJ...904...33L} {904, 33}

\bibitem[\protect\citeauthoryear{{Madau}}{{Madau}}{1995}]{Madau1995}
{Madau} P.,  1995, \mn@doi [\apj] {10.1086/175332}, \href
  {https://ui.adsabs.harvard.edu/abs/1995ApJ...441...18M} {441, 18}

\bibitem[\protect\citeauthoryear{{Meldorf}, {Palmese}  \& {Salim}}{{Meldorf}
  et~al.}{2023}]{Meldorf23}
{Meldorf} C.,  {Palmese} A.,   {Salim} S.,  2023, \mn@doi [arXiv e-prints]
  {10.48550/arXiv.2308.13974}, \href
  {https://ui.adsabs.harvard.edu/abs/2023arXiv230813974M} {p. arXiv:2308.13974}

\bibitem[\protect\citeauthoryear{{Pacifici} et~al.,}{{Pacifici}
  et~al.}{2015}]{Pacifici2015}
{Pacifici} C.,  et~al., 2015, \mn@doi [\mnras] {10.1093/mnras/stu2447}, \href
  {https://ui.adsabs.harvard.edu/abs/2015MNRAS.447..786P} {447, 786}

\bibitem[\protect\citeauthoryear{{Pacifici} et~al.,}{{Pacifici}
  et~al.}{2023}]{Pacifici23}
{Pacifici} C.,  et~al., 2023, \mn@doi [\apj] {10.3847/1538-4357/acacff}, \href
  {https://ui.adsabs.harvard.edu/abs/2023ApJ...944..141P} {944, 141}

\bibitem[\protect\citeauthoryear{{Peng}, {Ho}, {Impey}  \& {Rix}}{{Peng}
  et~al.}{2002}]{Peg2002}
{Peng} C.~Y.,  {Ho} L.~C.,  {Impey} C.~D.,   {Rix} H.-W.,  2002, \mn@doi [\aj]
  {10.1086/340952}, \href
  {https://ui.adsabs.harvard.edu/abs/2002AJ....124..266P} {124, 266}

\bibitem[\protect\citeauthoryear{Robitaille et~al.,}{Robitaille
  et~al.}{2013}]{robitaille_astropy:_2013}
Robitaille T.~P.,  et~al., 2013, \mn@doi [A\&A] {10.1051/0004-6361/201322068},
  558, A33

\bibitem[\protect\citeauthoryear{{Roper} et~al.,}{{Roper}
  et~al.}{2023}]{FLARES-IX}
{Roper} W.~J.,  et~al., 2023, arXiv e-prints, \href
  {https://ui.adsabs.harvard.edu/abs/2023arXiv230105228R} {p. arXiv:2301.05228}

\bibitem[\protect\citeauthoryear{Schaye et~al.,}{Schaye
  et~al.}{2015}]{schaye_eagle_2015}
Schaye J.,  et~al., 2015, \mn@doi [\mnras] {10.1093/mnras/stu2058}, 446, 521

\bibitem[\protect\citeauthoryear{{Snyder} et~al.,}{{Snyder}
  et~al.}{2015}]{Snyder15}
{Snyder} G.~F.,  et~al., 2015, \mn@doi [\mnras] {10.1093/mnras/stv2078}, \href
  {https://ui.adsabs.harvard.edu/abs/2015MNRAS.454.1886S} {454, 1886}

\bibitem[\protect\citeauthoryear{{Snyder}, {Pe{\~n}a}, {Yung}, {Rose},
  {Kartaltepe}  \& {Ferguson}}{{Snyder} et~al.}{2023}]{Snyder23}
{Snyder} G.~F.,  {Pe{\~n}a} T.,  {Yung} L.~Y.~A.,  {Rose} C.,  {Kartaltepe} J.,
    {Ferguson} H.,  2023, \mn@doi [\mnras] {10.1093/mnras/stac3397}, \href
  {https://ui.adsabs.harvard.edu/abs/2023MNRAS.518.6318S} {518, 6318}

\bibitem[\protect\citeauthoryear{{Somerville}, {Popping}  \&
  {Trager}}{{Somerville} et~al.}{2015}]{Somerville2015}
{Somerville} R.~S.,  {Popping} G.,   {Trager} S.~C.,  2015, \mn@doi [\mnras]
  {10.1093/mnras/stv1877}, \href
  {https://ui.adsabs.harvard.edu/abs/2015MNRAS.453.4337S} {453, 4337}

\bibitem[\protect\citeauthoryear{{Somerville} et~al.,}{{Somerville}
  et~al.}{2021}]{Somerville2021}
{Somerville} R.~S.,  et~al., 2021, \mn@doi [\mnras] {10.1093/mnras/stab231},
  \href {https://ui.adsabs.harvard.edu/abs/2021MNRAS.502.4858S} {502, 4858}

\bibitem[\protect\citeauthoryear{{Stanway} \& {Eldridge}}{{Stanway} \&
  {Eldridge}}{2018}]{BPASS2.2.1}
{Stanway} E.~R.,  {Eldridge} J.~J.,  2018, \mn@doi [\mnras]
  {10.1093/mnras/sty1353}, \href
  {https://ui.adsabs.harvard.edu/abs/2018MNRAS.479...75S} {479, 75}

\bibitem[\protect\citeauthoryear{{Vijayan}, {Lovell}, {Wilkins}, {Thomas},
  {Barnes}, {Irodotou}, {Kuusisto}  \& {Roper}}{{Vijayan}
  et~al.}{2021}]{FLARES-II}
{Vijayan} A.~P.,  {Lovell} C.~C.,  {Wilkins} S.~M.,  {Thomas} P.~A.,  {Barnes}
  D.~J.,  {Irodotou} D.,  {Kuusisto} J.,   {Roper} W.~J.,  2021, \mn@doi
  [\mnras] {10.1093/mnras/staa3715}, \href
  {https://ui.adsabs.harvard.edu/abs/2021MNRAS.501.3289V} {501, 3289}

\bibitem[\protect\citeauthoryear{{Villaume}, {Conroy}  \& {Johnson}}{{Villaume}
  et~al.}{2015}]{Villaume2015}
{Villaume} A.,  {Conroy} C.,   {Johnson} B.~D.,  2015, \mn@doi [\apj]
  {10.1088/0004-637X/806/1/82}, \href
  {https://ui.adsabs.harvard.edu/abs/2015ApJ...806...82V} {806, 82}

\bibitem[\protect\citeauthoryear{{Virtanen} et~al.,}{{Virtanen}
  et~al.}{2020}]{2020SciPy-NMeth}
{Virtanen} P.,  et~al., 2020, \mn@doi [Nature Methods]
  {https://doi.org/10.1038/s41592-019-0686-2}, \href {https://rdcu.be/b08Wh}
  {17, 261}

\bibitem[\protect\citeauthoryear{{Wilkins} et~al.,}{{Wilkins}
  et~al.}{2013}]{Wilkins13}
{Wilkins} S.~M.,  et~al., 2013, \mn@doi [\mnras] {10.1093/mnras/stt1471}, \href
  {https://ui.adsabs.harvard.edu/abs/2013MNRAS.435.2885W} {435, 2885}

\bibitem[\protect\citeauthoryear{{Wilkins}, {Feng}, {Di-Matteo}, {Croft},
  {Stanway}, {Bunker}, {Waters}  \& {Lovell}}{{Wilkins}
  et~al.}{2016}]{Wilkins16}
{Wilkins} S.~M.,  {Feng} Y.,  {Di-Matteo} T.,  {Croft} R.,  {Stanway} E.~R.,
  {Bunker} A.,  {Waters} D.,   {Lovell} C.,  2016, \mn@doi [\mnras]
  {10.1093/mnras/stw1154}, \href
  {https://ui.adsabs.harvard.edu/abs/2016MNRAS.460.3170W} {460, 3170}

\bibitem[\protect\citeauthoryear{{Wilkins} et~al.,}{{Wilkins}
  et~al.}{2022}]{FLARES-VI}
{Wilkins} S.~M.,  et~al., 2022, \mn@doi [\mnras] {10.1093/mnras/stac2548},
  \href {https://ui.adsabs.harvard.edu/abs/2022MNRAS.517.3227W} {517, 3227}

\bibitem[\protect\citeauthoryear{{Williams} et~al.,}{{Williams}
  et~al.}{2018}]{Williams2018}
{Williams} C.~C.,  et~al., 2018, \mn@doi [\apjs] {10.3847/1538-4365/aabcbb},
  \href {https://ui.adsabs.harvard.edu/abs/2018ApJS..236...33W} {236, 33}

\bibitem[\protect\citeauthoryear{{Yung}, {Somerville}, {Finkelstein}, {Popping}
   \& {Dav{\'e}}}{{Yung} et~al.}{2019a}]{Yung2019}
{Yung} L.~Y.~A.,  {Somerville} R.~S.,  {Finkelstein} S.~L.,  {Popping} G.,
  {Dav{\'e}} R.,  2019a, \mn@doi [\mnras] {10.1093/mnras/sty3241}, \href
  {https://ui.adsabs.harvard.edu/abs/2019MNRAS.483.2983Y} {483, 2983}

\bibitem[\protect\citeauthoryear{{Yung}, {Somerville}, {Popping},
  {Finkelstein}, {Ferguson}  \& {Dav{\'e}}}{{Yung} et~al.}{2019b}]{Yung2019b}
{Yung} L.~Y.~A.,  {Somerville} R.~S.,  {Popping} G.,  {Finkelstein} S.~L.,
  {Ferguson} H.~C.,   {Dav{\'e}} R.,  2019b, \mn@doi [\mnras]
  {10.1093/mnras/stz2755}, \href
  {https://ui.adsabs.harvard.edu/abs/2019MNRAS.490.2855Y} {490, 2855}

\bibitem[\protect\citeauthoryear{{Yung}, {Somerville}, {Popping}  \&
  {Finkelstein}}{{Yung} et~al.}{2020a}]{Yung2020a}
{Yung} L.~Y.~A.,  {Somerville} R.~S.,  {Popping} G.,   {Finkelstein} S.~L.,
  2020a, \mn@doi [\mnras] {10.1093/mnras/staa714}, \href
  {https://ui.adsabs.harvard.edu/abs/2020MNRAS.494.1002Y} {494, 1002}

\bibitem[\protect\citeauthoryear{{Yung}, {Somerville}, {Finkelstein},
  {Popping}, {Dav{\'e}}, {Venkatesan}, {Behroozi}  \& {Ferguson}}{{Yung}
  et~al.}{2020b}]{Yung2020b}
{Yung} L.~Y.~A.,  {Somerville} R.~S.,  {Finkelstein} S.~L.,  {Popping} G.,
  {Dav{\'e}} R.,  {Venkatesan} A.,  {Behroozi} P.,   {Ferguson} H.~C.,  2020b,
  \mn@doi [\mnras] {10.1093/mnras/staa1800}, \href
  {https://ui.adsabs.harvard.edu/abs/2020MNRAS.496.4574Y} {496, 4574}

\bibitem[\protect\citeauthoryear{{Yung}, {Somerville}, {Finkelstein},
  {Hirschmann}, {Dav{\'e}}, {Popping}, {Gardner}  \& {Venkatesan}}{{Yung}
  et~al.}{2021}]{Yung2021}
{Yung} L.~Y.~A.,  {Somerville} R.~S.,  {Finkelstein} S.~L.,  {Hirschmann} M.,
  {Dav{\'e}} R.,  {Popping} G.,  {Gardner} J.~P.,   {Venkatesan} A.,  2021,
  \mn@doi [\mnras] {10.1093/mnras/stab2761}, \href
  {https://ui.adsabs.harvard.edu/abs/2021MNRAS.508.2706Y} {508, 2706}

\bibitem[\protect\citeauthoryear{{Yung} et~al.,}{{Yung}
  et~al.}{2022}]{Yung2022}
{Yung} L.~Y.~A.,  et~al., 2022, \mn@doi [\mnras] {10.1093/mnras/stac2139},
  \href {https://ui.adsabs.harvard.edu/abs/2022MNRAS.515.5416Y} {515, 5416}

\bibitem[\protect\citeauthoryear{{Yung} et~al.,}{{Yung}
  et~al.}{2023}]{Yung2023}
{Yung} L.~Y.~A.,  et~al., 2023, \mn@doi [\mnras] {10.1093/mnras/stac3595},
  \href {https://ui.adsabs.harvard.edu/abs/2023MNRAS.519.1578Y} {519, 1578}

\makeatother
\end{thebibliography}





\bsp	
\label{lastpage}
\end{document}